\documentclass[runningheads]{llncs}
\usepackage{graphicx}
\begin{document}
\title{RowHammer and Beyond}
%
%
\author{Onur Mutlu}
%

%
\institute{ETH Z{\"u}rich and Carnegie Mellon University\\ 
\email{onur.mutlu@inf.ethz.ch}}
\maketitle              

\begin{abstract}
We will discuss the RowHammer problem in DRAM, which is a prime (and
likely the first) example of how a circuit-level failure mechanism in
Dynamic Random Access Memory (DRAM) can cause a practical and
widespread system security vulnerability.  RowHammer is the phenomenon
that repeatedly accessing a row in a modern DRAM chip predictably
causes errors in physically-adjacent rows. It is caused by a hardware
failure mechanism called read disturb errors.  Building on our initial
fundamental work that appeared at ISCA 2014, Google Project Zero
demonstrated that this hardware phenomenon can be exploited by
user-level programs to gain kernel privileges. Many other recent works
demonstrated other attacks exploiting RowHammer, including remote
takeover of a server vulnerable to RowHammer. We will analyze the root
causes of the problem and examine solution directions. We will also
discuss what other problems may be lurking in DRAM and other types of
memories, e.g., NAND flash and Phase Change Memory, which can
potentially threaten the foundations of reliable and secure systems,
as the memory technologies scale to higher densities.
\end{abstract}

\section{Summary}
  
As memory scales down to smaller technology nodes, new failure
mechanisms emerge that threaten its correct
operation~\cite{mutlu-imw13,onur-date17}. If such failures are not
anticipated and corrected, they can not only degrade system
reliability and availability but also, even more importantly, open up
new security vulnerabilities: a malicious attacker can exploit the
exposed failure mechanism to take over an entire system. As such, new
failure mechanisms in memory can become practical and significant
threats to system security.

In this keynote talk, based on our ISCA 2014
paper~\cite{rowhammer-isca2014}, we introduce the RowHammer problem in
DRAM, which is a prime (and likely the first) example of a real
circuit-level failure mechanism that causes a practical and widespread
system security vulnerability. RowHammer, as it is now popularly
referred to, is the phenomenon that repeatedly accessing a row in a
modern DRAM chip causes bit flips in physically-adjacent rows at
consistently predictable bit locations. It is caused by a hardware
failure mechanism called {\em DRAM disturbance errors}, which is a
manifestation of circuit-level cell-to-cell interference in a scaled
memory technology. Specifically, when a DRAM row is opened (i.e.,
activated) and closed (i.e., precharged) repeatedly (i.e., {\em
  hammered}), enough times within a DRAM refresh interval, one or more
bits in physically-adjacent DRAM rows can be flipped to the wrong
value. Using an FPGA-based DRAM testing
infrastructure~\cite{dram-isca2013,softmc}, we tested 129 DRAM modules
manufactured by three major manufacturers in seven recent years
(2008--2014) and found that 110 of them exhibited RowHammer errors,
the earliest of which dates back to 2010. Our ISCA 2014
paper~\cite{rowhammer-isca2014} provides a detailed and rigorous
analysis of various characteristics of RowHammer, including its data
pattern dependence, repeatability of errors, relationship with leaky
cells, and various circuit-level causes of the phenomenon.
  
We demonstrate that a very simple user-level
program~\cite{rowhammer-isca2014,safari-rowhammer} can reliably and
consistently induce RowHammer errors in commodity AMD and Intel
systems using vulnerable DRAM modules. We released the source code of
this program~\cite{safari-rowhammer}, which Google Project Zero later
enhanced~\cite{google-rowhammer-test}. Using our user-level RowHammer
program, we showed that both read and write accesses to memory can
induce bit flips, all of which occur in rows other than the one that
is being accessed. Since different DRAM rows are mapped to different
software pages, our user-level program could reliably corrupt specific
bits in pages belonging to other programs. As a result, RowHammer
errors can be exploited by a malicious program to breach memory
protection and compromise the system. In fact, we hypothesized, in our
ISCA 2014 paper, that our user-level program, with some engineering
effort, could be developed into a {\em disturbance attack} that
injects errors into other programs, crashes the system, or hijacks
control of the system.

RowHammer exposes a {\em security threat} since it leads to a serious
breach of memory isolation: an access to one memory row (e.g., an OS
page) predictably modifies the data stored in another row (e.g.,
another OS page). Malicious software, which we call {\em disturbance
  attacks}~\cite{rowhammer-isca2014}, or {\em RowHammer attacks}, can
be written to take advantage of these disturbance errors to take over
an entire system. Inspired by our ISCA 2014 paper's fundamental
findings, researchers from Google Project Zero demonstrated in 2015
that RowHammer can be effectively exploited by user-level programs to
gain kernel privileges on real
systems~\cite{google-project-zero,google-rh-blackhat}. Tens of other
works since then demonstrated other attacks exploiting
RowHammer. These include remote takeover of a server vulnerable to
RowHammer via JavaScript code execution~\cite{rowhammer-js}, takeover
of a victim virtual machine by another virtual machine running on the
same system~\cite{flip-feng-shui}, takeover of a mobile device by a
malicious user-level application that requires no
permissions~\cite{drammer}, takeover of a mobile system by triggering
RowHammer using the WebGL interface on a mobile GPU~\cite{glitch-vu},
takeover of a remote system by triggering RowHammer through the Remote
Direct Memory Access (RDMA) protocol~\cite{throwhammer,nethammer}, and
various other attacks
(e.g.,~\cite{cloudflops,dedup-est-machina,anotherflip,qiao2016new,bhattacharya2016curious,jang2017sgx,poddebniak2018attacking,aga2017good,tatar2018defeating,pessl2016drama}). Thus,
RowHammer has widespread and profound real implications on system
security, as it destroys memory isolation on top of which modern
system security principles are built.

We provide a wide variety of solutions, both {\em immediate} and {\em
  longer-term}, to RowHammer, starting from our ISCA 2014
paper~\cite{rowhammer-isca2014}. A popular {\em immediate} solution we
describe and analyze, is to increase the refresh rate of memory such
that the probability of inducing a RowHammer error before DRAM cells
get refreshed is reduced. Several major system manufacturers have
adopted this solution and released security patches that increased
DRAM refresh rates (e.g.,~\cite{rh-apple,rh-hp,rh-lenovo,rh-cisco}) in
memory controllers deployed in the field. While this solution is
practical and effective in reducing the vulnerability, assuming the
refresh rate is increased enough to avoid the vulnerability, it has
the significant drawbacks of increasing energy/power consumption,
reducing system performance, and degrading quality of service
experienced by user programs. Our paper shows that the refresh rate
needs to be increased by 7X if we want to eliminate {\em every single}
RowHammer-induced error we saw in our tests of 129 DRAM modules. Since
DRAM refresh is already a significant
burden~\cite{raidr,dram-isca2013,darp-hpca2014,kang-memforum2014,samira-sigmetrics14,avatar-dsn15,khan-dsn16,patel2017reach,vrl-dram}
on energy, performance, and QoS, increasing it by any significant
amount would only exacerbate the problem. Yet, increased refresh rate
is likely the most practical {\em immediate} solution to RowHammer
that can protect vulnerable chips that are already deployed in the
field.


After describing and analyzing six solutions to RowHammer, our ISCA
2014 paper shows that the long-term solution to RowHammer can actually
be simple and low cost. We introduce a new idea, called {\em PARA
  (Probabilistic Adjacent Row Activation)}: when the memory controller
closes a row (after it was activated), with a very low probability, it
refreshes the adjacent rows. The probability value is a parameter
determined by the system designer or provided programmatically, if
needed, to trade off between performance overhead and vulnerability
protection guarantees. We show that this solution is very effective:
it eliminates the RowHammer vulnerability, providing much higher
reliability guarantees than modern hard disks provide today, while
requiring no storage cost and having negligible performance and energy
overheads~\cite{rowhammer-isca2014}. Variants of this solution are
currently being adopted in DRAM chips and memory
controllers~\cite{x210-github,x210-rh-ss}.

The RowHammer problem leads to a new mindset that has enabled a renewed
interest in hardware security research: real memory chips are
vulnerable, in a simple and widespread manner, and this causes real
security problems.  We believe the RowHammer problem will worsen over
time since DRAM cells are getting closer to each other with technology
scaling. Other similar vulnerabilities may also be lurking in DRAM and
other types of memories, e.g., NAND flash memory or Phase Change
Memory, that can potentially threaten the foundations of secure
systems~\cite{onur-date17}. Our work advocates a principled
system-memory co-design approach to memory reliability and security
research that can enable us to better anticipate and prevent such
vulnerabilities.

%
%
%

\section{Significance, Impact and the Future}
  
RowHammer has spurred significant amount of research and industry
attention since its publication in 2014. Our ISCA 2014
paper~\cite{rowhammer-isca2014} is the first to experimentally and scientifically 
demonstrate the RowHammer vulnerability, its characteristics, and its prevalence in
real DRAM chips. RowHammer is a prime (and likely the first) example
of a hardware failure mechanism that causes a practical and widespread
system security vulnerability. Thus, the implications of RowHammer and
our ISCA 2014 paper on systems security is tremendous, both in the
short term and the long term: it is the first work we know of that
shows that a real reliability problem in one of the ubiquitous
general-purpose hardware components (DRAM chips) can cause practical
and widespread system security vulnerabilities.

Since its publication in 2014, RowHammer has already had significant
real-world impact on both industry and academia in at least four
directions. These directions will continue to exert long-term impact
for RowHammer, as memory cells continue to get closer to each other
while the technology scaling of memory continues.

First, our work has inspired many researchers to exploit RowHammer to
devise new attacks. As mentioned earlier, tens of papers were written
in top security venues that demonstrate various practical attacks
exploiting RowHammer
(e.g.,~\cite{cloudflops,dedup-est-machina,anotherflip,qiao2016new,bhattacharya2016curious,jang2017sgx,aga2017good,pessl2016drama,rowhammer-js,flip-feng-shui,drammer,glitch-vu}). These
attacks started with Google Project Zero's first work in
2015~\cite{google-project-zero,google-rh-blackhat} and they continue
to this date, with the latest ones that we know of being published in
Summer
2018~\cite{poddebniak2018attacking,nethammer,throwhammer,tatar2018defeating}. We
believe there is a lot more to come in this direction: as systems
security researchers understand more about RowHammer, and as the
RowHammer phenomenon continues to fundamentally affect memory chips
due to technology scaling problems~\cite{onur-date17}, researchers and
practitioners will develop different types of attacks to exploit
RowHammer in various contexts and in many more creative ways. Some
recent reports suggest that new-generation DDR4 DRAM chips are
vulnerable to RowHammer~\cite{rowhammer-thirdio,pessl2016drama,aga2017good,aichinger2015ddr}, so the fundamental security research
on RowHammer is likely to continue into the future.

Second, due to its prevalence in real DRAM chips, as demonstrated in
our ISCA 2014 paper, RowHammer has become a popular
phenomenon~\cite{rowhammer-wikipedia,rh-discuss,rh-twitter,rh-zdnet1,rowhammer-thirdio,google-rh-blackhat,rh-passmark,rh-futureplus,arstechnica_ddr4-rh},
which, in turn, has helped make hardware security even more "mainstream" in popular media and
the broader security community. It showed that hardware reliability 
problems can be very serious security threats that
have to be defended against. A well-read article from the Wired
magazine, all about RowHammer, is entitled "Forget Software -- Now
Hackers are Exploiting Physics!"~\cite{wired-rh}, indicating the shift
of mindset towards very low-level hardware security vulnerabilities in the popular
mainstream security community. Many other popular articles in press
have been written about RowHammer, many of which pointing to the
our ISCA 2014 work~\cite{rowhammer-isca2014} as the first demonstration and scientific analysis of the RowHammer problem. 
Showing that hardware reliability problems can be serious security threats and pulling them to the popular discussion
space, and thus influencing the mainstream discourse, creates a very long
term impact for the RowHammer problem.

Third, our work inspired many solution and mitigation techniques for
RowHammer from both researchers and industry practitioners. {\em
  Apple} publicly mentioned, in their critical security release for
RowHammer, that they increased the memory refresh rates due to the
"original research by Yoongu Kim et
al. (2014)"~\cite{rh-apple}. Memtest86 program was updated, including
a RowHammer test, acknowledging our ISCA 2014
paper~\cite{rh-passmark}. Many academic works developed solutions to
RowHammer, working from our original research
(e.g.,~\cite{anvil,moin-rowhammer,anotherflip,seyedzadeh2017counter,brasser2016can,irazoqui2016mascat,son2017making,gomez2016dram,van2018guardion,lee2018twice}). Multiple
industrial solutions (e.g.,~\cite{x210-github,x210-rh-ss}) were
inspired by our new solution to RowHammer, Probabilistic Adjacent Row
Activation (PARA). We believe such solutions will continue to be generated in both academia and industry, extending RowHammer's impact into the very long term.

Fourth, and perhaps most importantly, RowHammer enabled a shift of
mindset among mainstream security researchers: general-purpose
hardware is fallible (in a very widespread manner) and its problems
are actually exploitable. This shift of mindset enabled many systems
security researchers to examine hardware in more depth and understand
its inner workings and vulnerabilities better. We believe it is no
coincidence that two of the groups that concurrently discovered the
Meltdown~\cite{lipp2018meltdown} and Spectre~\cite{kocher2018spectre}
vulnerabilities (Google Project Zero and TU Graz InfoSec) have heavily
worked on RowHammer attacks before. We believe this shift in mindset,
enabled in good part by the existence and prevalence of RowHammer,
will continue to be very be important for discovering and solving
other potential vulnerabilities that may appear as a result of both
technology scaling and hardware design.

\section{Other Potential Vulnerabilities}

We believe that, as memory technologies scale to higher densities,
other problems may start appearing (or may already be going unnoticed)
that can potentially threaten the foundations of secure systems. There
have been recent large-scale field studies a well as small-scale
controlled studies of real memory errors on real devices and systems,
showing that both DRAM and NAND flash memory technologies are becoming
less reliable~\cite{superfri14,justin-memerrors-dsn15,dram-field-analysis2,dram-field-analysis3,dram-field-analysis4,justin-flash-sigmetrics15,flash-field-analysis2,cai2017errors,cai2017error,luo2018improving,luo2018heatwatch,cai-date12,cai-hpca15,mutlu-imw13,patel2017reach,onur-date17}. As detailed experimental
analyses of real DRAM and NAND flash chips show, both technologies are
becoming much more vulnerable to cell-to-cell interference
effects~\cite{superfri14,rowhammer-isca2014,cai-dsn15,cai-sigmetrics14,cai-iccd13,cai-date12,cai-date13,flash-fms-talk,yixin-jsac16,cai-hpca17,cai2017errors,cai2017error,mutlu-imw13,onur-date17}, data retention is becoming significantly more
difficult in both technologies~\cite{raidr,samira-sigmetrics14,dram-isca2013,khan-dsn16,avatar-dsn15,darp-hpca2014,kang-memforum2014,mandelman-jrd02,cai-hpca15,cai-iccd12,warm-msst15,cai-date12,cai-date13,cai-itj2013,flash-fms-talk,memcon-cal16,cai2017errors,cai2017error,luo2018improving,luo2018heatwatch,superfri14,mutlu-imw13}, and error variation
within and across chips is increasingly prominent~\cite{dram-isca2013,aldram,kevinchang-sigmetrics16,dram-process-variation-3,cai-date12,cai-date13,lee2017design,kim2018solar,kim2018dram,kim2019drange}.  Emerging memory technologies~\cite{mutlu-imw13,meza-weed13}, such as Phase-Change Memory~\cite{pcm-isca09,zhou-isca09,moin-isca09,moin-micro09,wong-pcm,raoux-pcm,pcm-ieeemicro10,pcm-cacm10,justin-taco14,rbla},
STT-MRAM~\cite{chen-ieeetmag10,kultursay-ispass13}, and
RRAM/ReRAM/
memristors~\cite{wong-rram} are likely to exhibit similar
and perhaps even more exacerbated reliability issues. We believe, if
not carefully accounted for and corrected, these reliability problems
may surface as security problems as well, as in the case of RowHammer,
especially if the technology is employed as part of the main memory
system that is directly exposed to user-level programs. We believe
future work examining these vulnerabilities, among others, is
promising for both fixing the vulnerabilities and enabling the
effective scaling of memory technology.

\section*{Acknowledgments}

This short paper and the associated keynote talk are heavily based on
two previous papers we have written on RowHammer, one that first
scientifically introduced and analyzed the phenomenon in ISCA 2014~\cite{rowhammer-isca2014} and
the other that provides an analysis and future outlook on
RowHammer~\cite{onur-date17}. They are a result of the research done
together with many students and collaborators over the course of the
past 7-8 years. In particular, three PhD theses have shaped the
understanding that led to this work. These are Yoongu Kim's thesis
entitled ``Architectural Techniques to Enhance DRAM
Scaling''~\cite{yoongu-thesis}, Yu Cai's thesis entitled ``NAND Flash
Memory: Characterization, Analysis, Modeling and
Mechanisms''~\cite{yucai-thesis} and his continued follow-on work
after his thesis, summarized in~\cite{cai2017errors,cai2017error}, and
Donghyuk Lee's thesis entitled ``Reducing DRAM Latency at Low Cost by
Exploiting Heterogeneity''~\cite{donghyuk-thesis-arxiv16}. We also
acknowledge various funding agencies (NSF, SRC, ISTC, CyLab) and
industrial partners (AliBaba, AMD, Google, Facebook, HP Labs, Huawei,
IBM, Intel, Microsoft, Nvidia, Oracle, Qualcomm, Rambus, Samsung,
Seagate, VMware) who have supported the presented and other related
work in my group generously over the years. The first version of this
talk was delivered at a CMU CyLab Partners Conference in September
2015. Another version of the talk was delivered as part of an Invited
Session at DAC 2016, with a collaborative accompanying paper entitled
``Who Is the Major Threat to Tomorrow’s Security? You, the Hardware
Designer''~\cite{dac-invited-paper16}. The most recent version is the
invited talk given at the Top Picks in Hardware and Embedded Security
workshop, co-located with ICCAD 2018~\cite{topinhes-workshop-url},
where RowHammer was selected as a Top Pick among hardware and embedded
security papers published between 2012-2017. I would like to also
thank Christina Giannoula for her help in preparing this manuscript.



\bibliographystyle{plain}
\bibliography{paper}

\begin{thebibliography}{100}

\bibitem{rh-discuss}
{RowHammer Discussion Group}.
\newblock \url{https://groups.google.com/forum/\#!forum /rowhammer-discuss}.

\bibitem{rh-twitter}
{RowHammer on Twitter}.
\newblock \url{https://twitter.com/search?q=rowhammer}.

\bibitem{safari-rowhammer}
{Rowhammer: Source Code for Testing the {Row Hammer} Error Mechanism in {DRAM}
  Devices.}
\newblock \url{https://github.com/CMU-SAFARI/rowhammer}.

\bibitem{google-rowhammer-test}
{Test DRAM for Bit Flips Caused by the RowHammer Problem}.
\newblock \url{https://github.com/google/rowhammer-test}.

\bibitem{x210-github}
{ThinkPad X210 BIOS Debugging}.
\newblock \url{https://github.com/tadfisher/x210-bios}.

\bibitem{x210-rh-ss}
{Tweet about RowHammer Mitigation on x210}.
\newblock \url{https://twitter.com/isislovecruft/status/1021939922754723841}.

\bibitem{topinhes-workshop-url}
{Top Picks in Hardware and Embedded Security - Workshop Collocated with ICCAD
  2018}.
\newblock \url{https://wp.nyu.edu/toppicksinhardwaresecurity/}, 2017.

\bibitem{aga2017good}
Misiker~Tadesse Aga, Zelalem~Birhanu Aweke, and Todd Austin.
\newblock {When Good Protections go Bad: Exploiting anti-DoS Measures to
  Accelerate Rowhammer Attacks}.
\newblock In {\em HOST}, 2017.

\bibitem{rh-futureplus}
Barbara Aichinger.
\newblock {The Known Failure Mechanism in DDR3 Memory referred to as Row
  Hammer}.
\newblock
  \url{http://ddrdetective.com/files/6414/1036/5710/The\_Known\_Failure\_Mechanism\_
  in\_DDR3\_memory\_referred\_to\_as\_Row\_Hammer.pdf}, {September} 2014.

\bibitem{aichinger2015ddr}
Barbara Aichinger.
\newblock {DDR Memory Errors Caused by Row Hammer}.
\newblock In {\em HPEC}, 2015.

\bibitem{rh-apple}
{Apple Inc.}
\newblock {About the security content of Mac EFI Security Update 2015-001}.
\newblock \url{https://support.apple.com/en-us/HT204934}, {June} 2015.

\bibitem{anvil}
Zelalem~Birhanu Aweke et~al.
\newblock Anvil: Software-based protection against next-generation rowhammer
  attacks.
\newblock In {\em ASPLOS}, 2016.

\bibitem{bhattacharya2016curious}
Sarani Bhattacharya and Debdeep Mukhopadhyay.
\newblock {Curious Case of RowHammer: Flipping Secret Exponent Bits using
  Timing Analysis}.
\newblock In {\em CHES}, 2016.

\bibitem{dedup-est-machina}
E.~Bosman et~al.
\newblock {Dedup Est Machina: Memory Deduplication as an Advanced Exploitation
  Vector}.
\newblock {\em {S\&P}}, 2016.

\bibitem{brasser2016can}
Ferdinand Brasser, Lucas Davi, David Gens, Christopher Liebchen, and Ahmad-Reza
  Sadeghi.
\newblock {Can't Touch This: Practical and Generic Software-only Defenses
  Against RowHammer Attacks}.
\newblock {\em USENIX Sec.}, 2017.

\bibitem{dac-invited-paper16}
W.~Burleson et~al.
\newblock {Who Is the Major Threat to Tomorrow's Security? You, the Hardware
  Designer}.
\newblock {\em {DAC}}, 2016.

\bibitem{cai-date12}
Y.~Cai et~al.
\newblock {Error Patterns in {MLC NAND} Flash Memory: Measurement,
  Characterization, and Analysis}.
\newblock In {\em DATE}, 2012.

\bibitem{cai-iccd12}
Y.~Cai et~al.
\newblock {{Flash Correct-and-Refresh}: Retention-Aware Error Management for
  Increased Flash Memory Lifetime}.
\newblock In {\em ICCD}, 2012.

\bibitem{cai-itj2013}
Y.~Cai et~al.
\newblock {Error Analysis and Retention-Aware Error Management for NAND Flash
  Memory}.
\newblock {\em ITJ}, 2013.

\bibitem{cai-iccd13}
Y.~Cai et~al.
\newblock {Program Interference in {MLC NAND} Flash Memory: Characterization,
  Modeling, and Mitigation}.
\newblock In {\em ICCD}, 2013.

\bibitem{cai-date13}
Y.~Cai et~al.
\newblock {Threshold Voltage Distribution in {MLC NAND} Flash Memory:
  Characterization, Analysis and Modeling}.
\newblock In {\em DATE}, 2013.

\bibitem{cai-sigmetrics14}
Y.~Cai et~al.
\newblock {Neighbor-Cell Assisted Error Correction for {MLC NAND} Flash
  Memories}.
\newblock In {\em SIGMETRICS}, 2014.

\bibitem{cai-hpca17}
Y.~Cai et~al.
\newblock {Vulnerabilities in MLC NAND Flash Memory Programming: Experimental
  Analysis, Exploits, and Mitigation Techniques}.
\newblock In {\em HPCA}, 2017.

\bibitem{yucai-thesis}
Yu~Cai.
\newblock {\em {NAND flash memory: Characterization, Analysis, Modeling and
  Mechanisms}}.
\newblock PhD thesis, Carnegie Mellon University, 2012.

\bibitem{cai-hpca15}
Yu~Cai et~al.
\newblock {Data Retention in {MLC NAND} Flash Memory: Characterization,
  Optimization and Recovery}.
\newblock In {\em HPCA}, 2015.

\bibitem{cai-dsn15}
Yu~Cai et~al.
\newblock {Read Disturb Errors in MLC NAND Flash Memory: Characterization,
  Mitigation, and Recovery}.
\newblock In {\em DSN}, 2015.

\bibitem{cai2017error}
Yu~Cai, Saugata Ghose, Erich~F Haratsch, Yixin Luo, and Onur Mutlu.
\newblock {Error Characterization, Mitigation, and Recovery in
  Flash-memory-based Solid-state Drives}.
\newblock {\em Proceedings of the IEEE}, 2017.

\bibitem{cai2017errors}
Yu~Cai, Saugata Ghose, Erich~F Haratsch, Yixin Luo, and Onur Mutlu.
\newblock {Errors in Flash-Memory-Based Solid-State Drives: Analysis,
  Mitigation, and Recovery}.
\newblock {\em arXiv preprint arXiv:1711.11427}, 2017.

\bibitem{dram-process-variation-3}
Karthik Chandrasekar et~al.
\newblock {Exploiting Expendable Process-margins in DRAMs for Run-time
  Performance Optimization}.
\newblock In {\em DATE}, 2014.

\bibitem{kevinchang-sigmetrics16}
K.~Chang et~al.
\newblock {Understanding Latency Variation in Modern DRAM Chips: Experimental
  Characterization, Analysis, and Optimization}.
\newblock {\em SIGMETRICS}, 2016.

\bibitem{darp-hpca2014}
Kevin Chang et~al.
\newblock {Improving {DRAM} Performance by Parallelizing Refreshes with
  Accesses}.
\newblock In {\em HPCA}, 2014.

\bibitem{chen-ieeetmag10}
E.~Chen et~al.
\newblock {Advances and Future Prospects of Spin-Transfer Torque Random Access
  Memory}.
\newblock {\em IEEE Transactions on Magnetics}, 2010.

\bibitem{vrl-dram}
Anup Das et~al.
\newblock {VRL-DRAM: Improving DRAM Performance via Variable Refresh Latency}.
\newblock In {\em DAC}, 2018.

\bibitem{rh-cisco}
Troy Fridley and Omar Santos.
\newblock {Mitigations Available for the DRAM Row Hammer Vulnerability}.
\newblock
  \url{http://blogs.cisco.com/security/mitigations-available-for-the-dram-row-hammer-vulnerability},
  {March} 2015.

\bibitem{glitch-vu}
P.~Frigo et~al.
\newblock {Grand Pwning Unit: Accelerating Microarchitectural Attacks with the
  GPU}.
\newblock {\em {IEEE S\&P}}, 2018.

\bibitem{gomez2016dram}
Hector Gomez, Andres Amaya, and Elkim Roa.
\newblock {DRAM Row-hammer Attack Reduction using Dummy Cells}.
\newblock In {\em NORCAS}, 2016.

\bibitem{arstechnica_ddr4-rh}
Dan Goodin.
\newblock {Once thought safe, DDR4 memory shown to be vulnerable to Rowhammer}.
\newblock
  \url{https://arstechnica.com/information-technology/2016/03/once-thought-safe-ddr4-memory-shown-to-be-vulnerable-to-rowhammer/},
  2016.

\bibitem{wired-rh}
Andy Greenberg.
\newblock {Forget Software -- Now Hackers are Exploiting Physics}.
\newblock
  \url{https://www.wired.com/2016/08/new-form-hacking-breaks-ideas-computers-work/},
  2016.

\bibitem{anotherflip}
D.~Gruss et~al.
\newblock {Another Flip in the Wall of Rowhammer Defenses}.
\newblock {\em {IEEE S\&P}}, 2018.

\bibitem{rowhammer-js}
Daniel Gruss et~al.
\newblock Rowhammer.js: {A} remote software-induced fault attack in javascript.
\newblock {\em CoRR}, abs/1507.06955, 2015.

\bibitem{rh-zdnet1}
Robin Harris.
\newblock {Flipping DRAM bits - maliciously}.
\newblock \url{http://www.zdnet.com/article/flipping-dram-bits-maliciously/},
  {December} 2014.

\bibitem{softmc}
Hasan Hassan et~al.
\newblock {SoftMC: A Flexible and Practical Open-Source Infrastructure for
  Enabling Experimental DRAM Studies}.
\newblock In {\em HPCA}, 2017.

\bibitem{rh-hp}
{Hewlett-Packard Enterprise}.
\newblock {HP Moonshot Component Pack Version 2015.05.0}.
\newblock
  \url{http://h17007.www1.hp.com/us/en/enterprise/servers/products/moonshot/
  component-pack/index.aspx}, 2015.

\bibitem{irazoqui2016mascat}
Gorka Irazoqui, Thomas Eisenbarth, and Berk Sunar.
\newblock {MASCAT: Stopping Microarchitectural Attacks Before Execution}.
\newblock {\em IACR Cryptology ePrint Archive}, 2016.

\bibitem{jang2017sgx}
Yeongjin Jang, Jaehyuk Lee, Sangho Lee, and Taesoo Kim.
\newblock {SGX-Bomb: Locking Down the Processor via Rowhammer Attack}.
\newblock In {\em SysTEX}, 2017.

\bibitem{kang-memforum2014}
Uksong Kang et~al.
\newblock {Co-Architecting Controllers and {DRAM} to Enhance {DRAM} Process
  Scaling}.
\newblock In {\em {The Memory Forum}}, 2014.

\bibitem{samira-sigmetrics14}
Samira Khan et~al.
\newblock {The Efficacy of Error Mitigation Techniques for {DRAM} Retention
  Failures: A Comparative Experimental Study}.
\newblock {\em SIGMETRICS}, 2014.

\bibitem{memcon-cal16}
Samira Khan et~al.
\newblock {A Case for Memory Content-Based Detection and Mitigation of
  Data-Dependent Failures in DRAM}.
\newblock {\em {CAL}}, 2016.

\bibitem{khan-dsn16}
Samira Khan et~al.
\newblock {PARBOR: An Efficient System-Level Technique to Detect Data-Dependent
  Failures in DRAM}.
\newblock In {\em DSN}, 2016.

\bibitem{moin-rowhammer}
Dae-Hyun Kim et~al.
\newblock {Architectural Support for Mitigating Row Hammering in DRAM
  Memories}.
\newblock {\em {IEEE CAL}}, 2015.

\bibitem{kim2018solar}
Jeremie~S Kim, Minesh Patel, Hasan Hassan, and Onur Mutlu.
\newblock {Solar-DRAM: Reducing DRAM Access Latency by Exploiting the Variation
  in Local Bitlines}.
\newblock In {\em ICCD}, 2018.

\bibitem{kim2018dram}
Jeremie~S. Kim, Minesh Patel, Hasan Hassan, and Onur Mutlu.
\newblock {The DRAM Latency PUF: Quickly Evaluating Physical Unclonable
  Functions by Exploiting the Latency-Reliability Tradeoff in Modern Commodity
  DRAM Devices}.
\newblock In {\em HPCA}, 2018.

\bibitem{kim2019drange}
Jeremie~S. Kim, Minesh Patel, Hasan Hassan, Lois Orosa, and Onur Mutlu.
\newblock {D-RaNGe: Using Commodity DRAM Devices to Generate True Random
  Numbers with Low Latency and High Throughput}.
\newblock In {\em HPCA}, 2019.

\bibitem{yoongu-thesis}
Yoongu Kim.
\newblock {\em {Architectural Techniques to Enhance DRAM Scaling}}.
\newblock PhD thesis, Carnegie Mellon University, 2015.

\bibitem{rowhammer-isca2014}
Yoongu Kim et~al.
\newblock {Flipping Bits in Memory Without Accessing Them: An Experimental
  Study of {DRAM} Disturbance Errors}.
\newblock In {\em ISCA}, 2014.

\bibitem{kocher2018spectre}
Paul Kocher, Daniel Genkin, Daniel Gruss, Werner Haas, Mike Hamburg, Moritz
  Lipp, Stefan Mangard, Thomas Prescher, Michael Schwarz, and Yuval Yarom.
\newblock {Spectre Attacks: Exploiting Speculative Execution}.
\newblock {\em S\&P}, 2018.

\bibitem{kultursay-ispass13}
E.~Kultursay et~al.
\newblock Evaluating {STT-RAM} as an energy-efficient main memory alternative.
\newblock In {\em ISPASS}, 2013.

\bibitem{rowhammer-thirdio}
Mark Lanteigne.
\newblock {How Rowhammer Could Be Used to Exploit Weaknesses in Computer
  Hardware}.
\newblock \url{http://www.thirdio.com/rowhammer.pdf}, {March} 2016.

\bibitem{pcm-isca09}
B.~C. Lee et~al.
\newblock {Architecting Phase Change Memory as a Scalable {DRAM} Alternative}.
\newblock In {\em ISCA}, 2009.

\bibitem{pcm-cacm10}
B.~C. Lee et~al.
\newblock {Phase Change Memory Architecture and the Quest for Scalability}.
\newblock {\em {CACM}}, 2010.

\bibitem{pcm-ieeemicro10}
Benjamin~C. Lee et~al.
\newblock {Phase Change Technology and the Future of Main Memory}.
\newblock {\em IEEE Micro}, 2010.

\bibitem{donghyuk-thesis-arxiv16}
D.~Lee.
\newblock {Reducing DRAM Latency by Exploiting Heterogeneity}.
\newblock {\em {ArXiV}}, 2016.

\bibitem{aldram}
D.~Lee et~al.
\newblock {Adaptive-Latency {DRAM}: Optimizing {DRAM} Timing for the
  Common-Case}.
\newblock In {\em HPCA}, 2015.

\bibitem{lee2017design}
Donghyuk Lee, Samira Khan, Lavanya Subramanian, Saugata Ghose, Rachata
  Ausavarungnirun, Gennady Pekhimenko, Vivek Seshadri, and Onur Mutlu.
\newblock {Design-induced Latency Variation in Modern DRAM Chips:
  Characterization, Analysis, and Latency Reduction Mechanisms}.
\newblock {\em POMACS}, 2017.

\bibitem{lee2018twice}
Eojin Lee, Sukhan Lee, G~Edward Suh, and Jung~Ho Ahn.
\newblock {TWiCe: Time Window Counter Based Row Refresh to Prevent
  Row-Hammering}.
\newblock {\em CAL}, 2018.

\bibitem{rh-lenovo}
{Lenovo}.
\newblock {Row Hammer Privilege Escalation}.
\newblock \url{https://support.lenovo.com/us/en/product\_security/row\_hammer},
  {March} 2015.

\bibitem{nethammer}
M.~Lipp et~al.
\newblock {Nethammer: Inducing Rowhammer Faults through Network Requests}.
\newblock {\em {arxiv.org}}, 2018.

\bibitem{lipp2018meltdown}
Moritz Lipp, Michael Schwarz, Daniel Gruss, Thomas Prescher, Werner Haas,
  Anders Fogh, Jann Horn, Stefan Mangard, Paul Kocher, Daniel Genkin, et~al.
\newblock {Meltdown: Reading Kernel Memory from User Space}.
\newblock In {\em USENIX Security}, 2018.

\bibitem{raidr}
J.~Liu et~al.
\newblock {RAIDR}: Retention-aware intelligent {DRAM} refresh.
\newblock {\em ISCA}, 2012.

\bibitem{dram-isca2013}
J.~Liu et~al.
\newblock An experimental study of data retention behavior in modern {DRAM}
  devices: Implications for retention time profiling mechanisms.
\newblock {\em ISCA}, 2013.

\bibitem{warm-msst15}
Y.~Luo et~al.
\newblock {WARM: Improving NAND Flash Memory Lifetime with Write-hotness Aware
  Retention Management}.
\newblock {\em {MSST}}, 2015.

\bibitem{yixin-jsac16}
Yixin Luo et~al.
\newblock {Enabling Accurate and Practical Online Flash Channel Modeling for
  Modern MLC NAND Flash Memory}.
\newblock {\em {JSAC}}, 2016.

\bibitem{luo2018heatwatch}
Yixin Luo, Saugata Ghose, Yu~Cai, Erich~F Haratsch, and Onur Mutlu.
\newblock {HeatWatch: Improving 3D NAND Flash Memory Device Reliability by
  Exploiting Self-Recovery and Temperature Awareness}.
\newblock In {\em HPCA}, 2018.

\bibitem{luo2018improving}
Yixin Luo, Saugata Ghose, Yu~Cai, Erich~F Haratsch, and Onur Mutlu.
\newblock {Improving 3D NAND Flash Memory Lifetime by Tolerating Early
  Retention Loss and Process Variation}.
\newblock {\em POMACS}, 2018.

\bibitem{mandelman-jrd02}
J.~Mandelman et~al.
\newblock Challenges and future directions for the scaling of dynamic
  random-access memory ({DRAM}).
\newblock {\em {IBM Journal of Research and Development}}, 46, 2002.

\bibitem{meza-weed13}
J.~Meza et~al.
\newblock {A Case for Efficient Hardware-Software Cooperative Management of
  Storage and Memory}.
\newblock In {\em WEED}, 2013.

\bibitem{justin-flash-sigmetrics15}
J.~Meza et~al.
\newblock {A Large-Scale Study of Flash Memory Errors in the Field}.
\newblock In {\em SIGMETRICS}, 2015.

\bibitem{justin-memerrors-dsn15}
J.~Meza et~al.
\newblock {Revisiting Memory Errors in Large-Scale Production Data Centers:
  Analysis and Modeling of New Trends from the Field}.
\newblock {\em DSN}, 2015.

\bibitem{mutlu-imw13}
O.~Mutlu.
\newblock {Memory Scaling: A Systems Architecture Perspective}.
\newblock {\em IMW}, 2013.

\bibitem{onur-date17}
O.~Mutlu.
\newblock {The RowHammer Problem and Other Issues we may Face as Memory Becomes
  Denser}.
\newblock {\em {DATE}}, 2017.

\bibitem{flash-fms-talk}
Onur Mutlu.
\newblock {Error Analysis and Management for MLC NAND Flash Memory}.
\newblock In {\em {Flash Memory Summit}}, 2014.

\bibitem{superfri14}
Onur Mutlu and Lavanya Subramanian.
\newblock Research problems and opportunities in memory systems.
\newblock {\em SUPERFRI}, 2014.

\bibitem{rh-passmark}
{PassMark Software}.
\newblock {MemTest86: The Original Industry Standard Memory Diagnostic
  Utility}.
\newblock \url{http://www.memtest86.com/troubleshooting.htm}, 2015.

\bibitem{patel2017reach}
Minesh Patel, Jeremie~S Kim, and Onur Mutlu.
\newblock {The Reach Profiler (REAPER): Enabling the Mitigation of DRAM
  Retention Failures via Profiling at Aggressive Conditions}.
\newblock {\em ISCA}, 2017.

\bibitem{pessl2016drama}
Peter Pessl, Daniel Gruss, Cl{\'e}mentine Maurice, Michael Schwarz, and Stefan
  Mangard.
\newblock {DRAMA: Exploiting DRAM Addressing for Cross-CPU Attacks}.
\newblock In {\em USENIX Security}, 2016.

\bibitem{poddebniak2018attacking}
Damian Poddebniak, Juraj Somorovsky, Sebastian Schinzel, Manfred Lochter, and
  Paul R{\"o}sler.
\newblock {Attacking Deterministic Signature Schemes using Fault Attacks}.
\newblock In {\em EuroS\&P}, 2018.

\bibitem{qiao2016new}
Rui Qiao and Mark Seaborn.
\newblock {A New Approach for Rowhammer Attacks}.
\newblock In {\em HOST}, 2016.

\bibitem{moin-isca09}
M.~K. Qureshi et~al.
\newblock Scalable high performance main memory system using phase-change
  memory technology.
\newblock In {\em ISCA}, 2009.

\bibitem{avatar-dsn15}
M.~K. Qureshi et~al.
\newblock {AVATAR: A Variable-Retention-Time (VRT) Aware Refresh for DRAM
  Systems}.
\newblock In {\em DSN}, 2015.

\bibitem{moin-micro09}
Moinuddin~K. Qureshi et~al.
\newblock {Enhancing Lifetime and Security of Phase Change Memories via
  Start-Gap Wear Leveling}.
\newblock In {\em MICRO}, 2009.

\bibitem{raoux-pcm}
S.~Raoux et~al.
\newblock {Phase-change Random Access Memory: A Scalable Technology}.
\newblock {\em {IBM Journal of Research and Development}}, 2008.

\bibitem{flip-feng-shui}
K.~Razavi et~al.
\newblock {Flip Feng Shui: Hammering a Needle in the Software Stack}.
\newblock {\em {USENIX Security}}, 2016.

\bibitem{flash-field-analysis2}
Bianca Schroeder et~al.
\newblock {Flash Reliability in Production: The Expected and the Unexpected}.
\newblock In {\em {USENIX FAST}}, 2016.

\bibitem{google-project-zero}
Mark Seaborn and Thomas Dullien.
\newblock {Exploiting the DRAM Rowhammer Bug to Gain Kernel Privileges}.
\newblock
  \url{http://googleprojectzero.blogspot.com.tr/2015/03/exploiting-dram-rowhammer-\newline
bug-to-gain.html},
  2015.

\bibitem{google-rh-blackhat}
Mark Seaborn and Thomas Dullien.
\newblock {Exploiting the DRAM rowhammer bug to gain kernel privileges}.
\newblock {\em BlackHat}, 2016.

\bibitem{seyedzadeh2017counter}
Seyed~Mohammad Seyedzadeh, Alex~K Jones, and Rami Melhem.
\newblock {Counter-based Tree Structure for Row Hammering Mitigation in DRAM}.
\newblock {\em CAL}, 2017.

\bibitem{son2017making}
Mungyu Son, Hyunsun Park, Junwhan Ahn, and Sungjoo Yoo.
\newblock {Making DRAM Stronger Against Row Hammering}.
\newblock In {\em DAC}, 2017.

\bibitem{dram-field-analysis2}
Vilas Sridharan, Nathan DeBardeleben, Sean Blanchard, Kurt~B. Ferreira, Jon
  Stearley, John Shalf, and Sudhanva Gurumurthi.
\newblock {Memory Errors in Modern Systems: The Good, The Bad, and The Ugly}.
\newblock In {\em ASPLOS}, 2015.

\bibitem{dram-field-analysis3}
Vilas Sridharan and Dean Liberty.
\newblock {A Study of {DRAM} Failures in the Field}.
\newblock In {\em {SC}}, 2012.

\bibitem{dram-field-analysis4}
Vilas Sridharan, Jon Stearley, Nathan DeBardeleben, Sean Blanchard, and
  Sudhanva Gurumurthi.
\newblock {Feng Shui of Supercomputer Memory: Positional Effects in {DRAM} and
  {SRAM} Faults}.
\newblock In {\em SC}, 2013.

\bibitem{throwhammer}
A.~Tatar et~al.
\newblock {Throwhammer: Rowhammer Attacks over the Network and Defenses}.
\newblock {\em {USENIX ATC}}, 2018.

\bibitem{tatar2018defeating}
Andrei Tatar, Cristiano Giuffrida, Herbert Bos, and Kaveh Razavi.
\newblock {Defeating Software Mitigations Against Rowhammer: A Surgical
  Precision Hammer}.
\newblock In {\em RAID}, 2018.

\bibitem{drammer}
V.~van~der Veen et~al.
\newblock {Drammer: Deterministic Rowhammer Attacks on Mobile Platforms}.
\newblock {\em {CCS}}, 2016.

\bibitem{van2018guardion}
Victor van~der Veen, Martina Lindorfer, Yanick Fratantonio,
  Harikrishnan~Padmanabha Pillai, Giovanni Vigna, Christopher Kruegel, Herbert
  Bos, and Kaveh Razavi.
\newblock {GuardION: Practical Mitigation of DMA-Based Rowhammer Attacks on
  ARM}.
\newblock In {\em DIMVA}, 2018.

\bibitem{rowhammer-wikipedia}
Wikipedia.
\newblock {Row hammer}.
\newblock \url{https://en.wikipedia.org/wiki/Row\_hammer}.

\bibitem{wong-pcm}
H-S.~P. Wong et~al.
\newblock {Phase Change Memory}.
\newblock {\em Proceedings of the IEEE}, 2010.

\bibitem{wong-rram}
H-S.~P. Wong et~al.
\newblock {Metal-Oxide {RRAM}}.
\newblock In {\em Proceedings of the IEEE}, 2012.

\bibitem{cloudflops}
Y.~Xiao et~al.
\newblock {One Bit Flips, One Cloud Flops: Cross-VM Row Hammer Attacks and
  Privilege Escalation}.
\newblock {\em {USENIX Sec.}}, 2016.

\bibitem{rbla}
H.~Yoon et~al.
\newblock {Row Buffer Locality Aware Caching Policies for Hybrid Memories}.
\newblock In {\em ICCD}, 2012.

\bibitem{justin-taco14}
HanBin Yoon et~al.
\newblock {Efficient Data Mapping and Buffering Techniques for Multi-Level Cell
  Phase-Change Memories}.
\newblock {\em TACO}, 2014.

\bibitem{zhou-isca09}
Ping Zhou et~al.
\newblock {A Durable and Energy Efficient Main Memory using Phase Change Memory
  Technology}.
\newblock In {\em ISCA}, 2009.

\end{thebibliography}

\end{document}